\documentclass{article}
\usepackage{codefuse_tech_report}
\usepackage[colorlinks = true,
            linkcolor = blue,
            urlcolor  = blue,
            citecolor = blue,
            anchorcolor = blue]{hyperref}   
\usepackage{microtype}
\usepackage{xurl}
\usepackage{booktabs}
\usepackage{enumitem}
\usepackage{multicol}
\usepackage{CJKutf8}
\usepackage{siunitx}
\usepackage{floatflt}
\usepackage{graphicx}
\usepackage{wrapfig}
\usepackage{authblk}
\usepackage{lipsum}
\usepackage{algorithm}
\usepackage{algorithmicx}
\usepackage{algpseudocode}
\usepackage{multirow}
\usepackage{pifont}  
\usepackage{tabularx}
\usepackage{subcaption}  

\algnewcommand{\LeftComment}[1]{\Statex \(\triangleright\) #1}

\usepackage{array}
\usepackage{amsmath}
\usepackage{amssymb}
\usepackage{mathtools}
\usepackage{amsthm}

\usepackage[capitalize,noabbrev]{cleveref}
\usepackage{adjustbox} 
\usepackage{fancyvrb}    
\usepackage{tcolorbox}   

\theoremstyle{plain}

\theoremstyle{definition}

\theoremstyle{remark}

\colmfinalcopy

\usepackage[utf8]{inputenc}
\usepackage[T1]{fontenc}
\usepackage{caption} 
\usepackage{adjustbox} 
\usepackage{fontawesome5}

\usepackage{tcolorbox}
\tcbuselibrary{skins,breakable}

\tcbuselibrary{skins}

\usepackage{color}
\usepackage{xcolor}
\usepackage{soul} 

\definecolor{tred}{RGB}{251, 130, 132}
\definecolor{torange}{RGB}{247, 162, 116}
\definecolor{tyellow}{RGB}{251, 218, 140}
\definecolor{tgreen}{RGB}{127, 204, 181}
\definecolor{tblue}{RGB}{89, 177, 215}
\definecolor{insightblue}{RGB}{162, 210, 255}
\definecolor{questionred}{RGB}{255, 175, 204}

\newcolumntype{Y}{>{\centering\arraybackslash}X}

\newcommand{\rubrictable}[1]{
    \noindent
    \begin{tabularx}{\textwidth}{@{} l X @{}}
        \toprule
        \textbf{Score} & \textbf{Criteria} \\
        \midrule
        #1 \\
        \bottomrule
    \end{tabularx}
    \vspace{1em}
}

\newtcolorbox{inputbox}[2][]{
    colback=gray!5!white,      
    colframe=green!40!black,    
    fonttitle=\bfseries,        
    title=#2,                   
    arc=2mm,                    
    boxrule=0.5pt,              
    left=2mm, right=2mm, top=2mm, bottom=2mm, 
    enhanced,
    attach boxed title to top left={xshift=3mm, yshift=-2mm}, 
    boxed title style={colback=green!40!black}, 
    breakable,                           
    #1
}

\newtcolorbox{casebox}[2][]{
    colback=blue!3!white,                
    colframe=blue!50!black,              
    fonttitle=\bfseries,
    title=#2,
    arc=2mm,
    boxrule=0.5pt,
    left=2mm, right=2mm, top=2mm, bottom=2mm,
    enhanced,
    attach boxed title to top left={xshift=3mm, yshift=-2mm},
    boxed title style={colback=blue!50!black}, 
    breakable,                           
    #1
}

\newtcolorbox{outputbox}[2][]{
    colback=orange!4!white,              
    colframe=orange!60!black,            
    fonttitle=\bfseries,
    title=#2,
    arc=2mm,
    boxrule=0.5pt,
    left=2mm, right=2mm, top=2mm, bottom=2mm,
    enhanced,
    attach boxed title to top left={xshift=3mm, yshift=-2mm},
    boxed title style={colback=orange!60!black},
    breakable,                           
    #1
}

\usepackage{listings}
\title{EGSS: Entropy-guided Stepwise Scaling for Reliable Software Engineering}

\author{%
Chenhui Mao\thanks{Equal Contribution.}$^{\phantom{*}, 1}$
~~Yuanting Lei$^{*, 1}$
~~Zhixiang Wei$^{1}$ ~~Ming Liang$^{1}$ ~~Zhixiang Wang$^{1}$ ~~Jingxuan Xu$^{1}$ ~~Dajun Chen$^{1}$ ~~Wei Jiang$^{1}$ ~~Yong Li$^{1}$\thanks{Corresponding author.\\ \{maochenhui.maochen, leiyuanting.ly,weizhixiang.wzx,liangming.liang,tiejing.wzx,xujingxuan.xjx, \\ chendajun.cdj,jonny.jw,liyong.liy\}@antgroup.com}

\vspace{10pt}
$^1$ Ant Group \\
\vspace{10pt}
\hspace{-10pt}\faGithub ~\url{https://github.com/codefuse-ai/CodeFuse-Agent}\\
}

\colmfinalcopy 

\begin{document}

\maketitle

\begin{figure}[h!]
    \centering
    \includegraphics[width=0.5\linewidth]{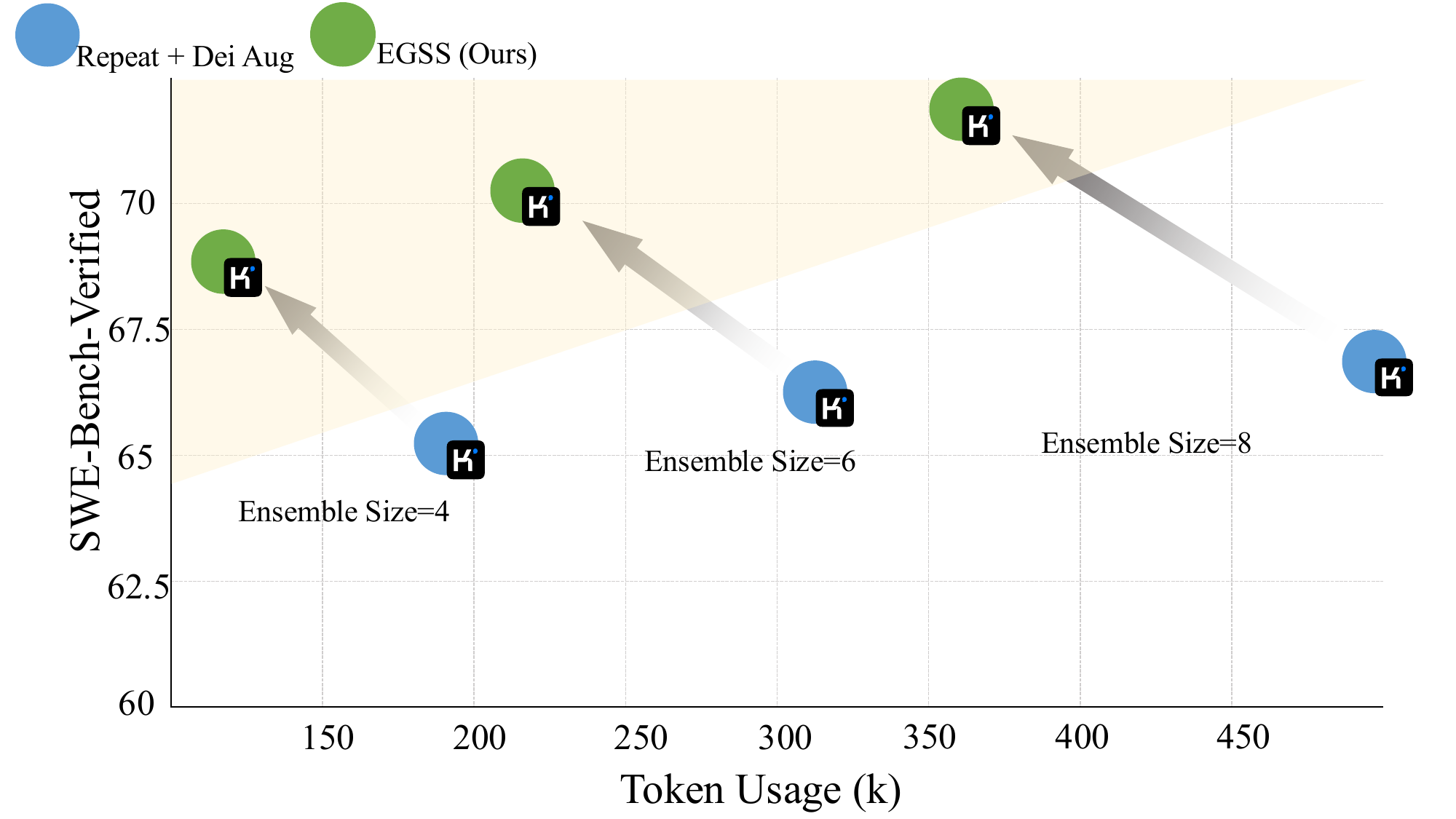}
    \caption{Performance and token usage of popular test-time scaling methods compared with entropy-guided stepwise scaling on SWE-Bench Verified, using {Kimi-K2-Instruct} as the base model.}
    \label{fig:trade_off}
\end{figure}

\begin{abstract}
Agentic Test-Time Scaling (TTS) has delivered state-of-the-art (SOTA) performance on complex software engineering tasks such as code generation and bug fixing. 
However, its practical adoption remains limited due to significant computational overhead, primarily driven by two key challenges: (1) the high cost associated with deploying excessively large ensembles, and (2) the lack of a reliable mechanism for selecting the optimal candidate solution—ultimately constraining the performance gains that can be realized.
To address these challenges, we propose \textbf{Entropy-Guided Stepwise Scaling (EGSS)}, a novel TTS framework that dynamically balances efficiency and effectiveness through entropy-guided adaptive search and robust test-suite augmentation.  
Extensive experiments on SWE-Bench-Verified demonstrate that EGSS consistently boosts performance by 5–10\% across all evaluated models. Specifically, it increases the resolved ratio of Kimi-K2-Intruct from \textbf{63.2\%} to \textbf{72.2\%}, and GLM-4.6 from \textbf{65.8\%} to \textbf{74.6\%}. Furthermore, when paired with GLM-4.6, EGSS achieves new state-of-the-art among open-source large language models. In addition to these accuracy improvements, EGSS reduces inference-time token usage by over \textbf{28\%} compared to existing TTS methods, achieving simultaneous gains in both effectiveness and computational efficiency.
\end{abstract}

\section{Introduction}
Autonomous Software Engineering (ASE) has broadened the application scope of large language models (LLMs) for code from generating simple functions to resolving complex, repository-level software issues. In contrast to snippet-level tasks~\cite{chen2021evaluating}, benchmarks such as SWE-bench~\cite{jimenez2023swe} require agents to operate within large-scale codebases, thereby necessitating long-horizon reasoning, proficient tool utilization, and a holistic understanding of cross-module dependencies~\cite{yang2024sweagent, wu2024introducing, bouzenia2024repairagent}. As task complexity increases, the expanding action space and the inherent uncertainty of debugging trajectories substantially impede effective decision-making, often resulting in diminished reliability.

Within this demanding landscape, Test-Time Scaling (TTS) has emerged as a pivotal paradigm, demonstrating substantial potential to recover performance on complex problems by strategically allocating additional inference-time compute to explore diverse reasoning trajectories~\cite{zeng2025satori, gao2025trae}. However, despite its promise, existing TTS strategies often entail considerable computational overhead and exhibit limited reliability. 
\textbf{First}, prevailing TTS approaches frequently suffer from pronounced computational redundancy and low exploration efficiency~\cite{li2025s}. They commonly adopt uniform search expansion or large-scale repetitive sampling without explicit inter-trajectory coordination, thereby expending resources on trivial operations and repeatedly traversing unproductive branches. Moreover, many frameworks~\cite{gandhi2025agents} employ static process-level guidance that does not adapt to the time-varying uncertainty inherent in complex tasks. This induces a resource--uncertainty mismatch, wherein computation is over-allocated to relatively straightforward instances while insufficient exploration is devoted to high-ambiguity decision points. 
\textbf{Second}, existing patch selection mechanisms often discard informative intermediate debugging signals~\cite{wang2025steca}. By disregarding execution traces, terminal-only ranking methods are susceptible to "consensus errors" and may favor brittle solutions that satisfy localized checks without addressing the underlying root cause. These limitations motivate the following core question:

{\centering
\textit{Can we design \textbf{principled} methods to make TTS both \textbf{efficient} and \textbf{reliable} within the context of ASE?}
}

To address these challenges, we propose Entropy-Guided Stepwise Scaling (EGSS), a principled, entropy-guided framework that jointly optimizes efficiency and reliability for ASE. 

\textbf{First}, to mitigate redundant exploration in long-horizon reasoning, EGSS introduces a criterion based on tool entropy~\cite{dong2025agentic} to identify critical decision points, at which an auxiliary judge is selectively invoked to conduct real-time evaluation and prune branching trajectories. By using entropy as a measure of decision ambiguity, this mechanism dynamically reallocates the inference budget toward steps where exploration is most informative, while systematically avoiding redundant or low-utility actions. As a result, EGSS substantially improves computational efficiency without compromising solution quality (see Figure~\ref{fig:trade_off} for the resulting trade-off between token consumption and task performance).

\textbf{Second}, to further improve the reliability of patch selection, EGSS constructs a comprehensive, multi-dimensional test suite by integrating debugging signals extracted from multiple agent trajectories with existing regression tests in the repository and newly identified edge cases. This enriched test suite is then used to systematically filter low-quality candidate patches. In addition, EGSS employs a multi-model ensemble voting mechanism to produce robust and consistent patch proposals. By incorporating entropy-guided trajectory selection, EGSS reformulates large-scale code exploration as a resource-aware, closed-loop optimization process, thereby enabling efficient and reliable execution of autonomous software engineering tasks.

Overall, our primary contributions are summarized as follows:
\begin{itemize} 
    \item We present a principled, entropy-guided scaling paradigm that improves inference efficiency by allocating inference budget to high-ambiguity decision points.
    \item We develop a robust patch selection pipeline that consolidates debugging signals into a high-coverage test suite, mitigating false positives and prioritizing robust candidate patches.
    \item We propose EGSS, which improves the performance of all evaluated models by 5--10\% on SWE-Bench-Verified, establishing a new state of the art among open-source methods, while reducing inference-time token consumption by 28\% relative to prior TTS approaches.
\end{itemize}
\section{Related Work}
\subsection{Scaling Paradigms and Search Efficiency}
Previous agentic frameworks~\cite{yang2024sweagent, xie2024autocoderover, wu2024introducing} established repository-level repair protocols but primarily relied on single, linear reasoning trajectories. To address this limitation, Test-Time Scaling has emerged as a prominent paradigm that improves performance by allocating additional inference-time computation to explore a broader solution space~\cite{ma2025thinking, li2025s}. While existing methods employ heuristic search~\cite{antoniades2024swe} or sampling in tool-interactive, non-deterministic environments~\cite{zainullina2025guided}, they often exhibit substantial redundancy under uniform scaling, expanding the search space regardless of step-level decision criticality. Moreover, although recent process-guided models~\cite{gandhi2025agents, li2025codeprm, sun2025freeprm} provide intermediate feedback, their reliance on static diagnostics~\cite{khalifa2025process} limits their ability to adaptively identify high-uncertainty decision points.

\subsection{Reliability in Patch Selection}
A key bottleneck in multi-trajectory test-time scaling is the limited reliability of final patch selection. Common strategies, ranging from majority voting~\cite{xia2024agentless} to multi-agent collaboration~\cite{zhang2024diversity}, can mitigate sampling-induced errors but remain vulnerable to consensus errors, where multiple trajectories converge on a plausible yet incorrect fix. More advanced selector agents~\cite{zeng2025satori, gao2025trae} and trajectory calibration methods~\cite{wang2025steca} attempt to rank candidates, but they are still predominantly terminal-stage (post-hoc) selectors. Consequently, they treat trajectories as isolated candidates and discard rich intermediate debugging signals generated during the reasoning process~\cite{wang2024executable}. Even with specialized verifiers~\cite{pan2024training, kong2025contrastrepair}, the absence of cross-trajectory signal synthesis can lead to brittle patches.
\section{Preliminary}
In this section, we conduct a systematic analysis of an agent instantiated with Kimi-K2-Instruct and evaluate its performance on SWE-Bench tasks~\cite{jimenez2023swe}. The results reveal several salient empirical phenomena, which provide direct theoretical and practical motivation for the design of the proposed Entropy-guided Stepwise Scaling architecture.

\subsection{Tool Use Entropy Heterogeneity}
\begin{figure}[t]
  \centering
  \includegraphics[width=0.5\linewidth]{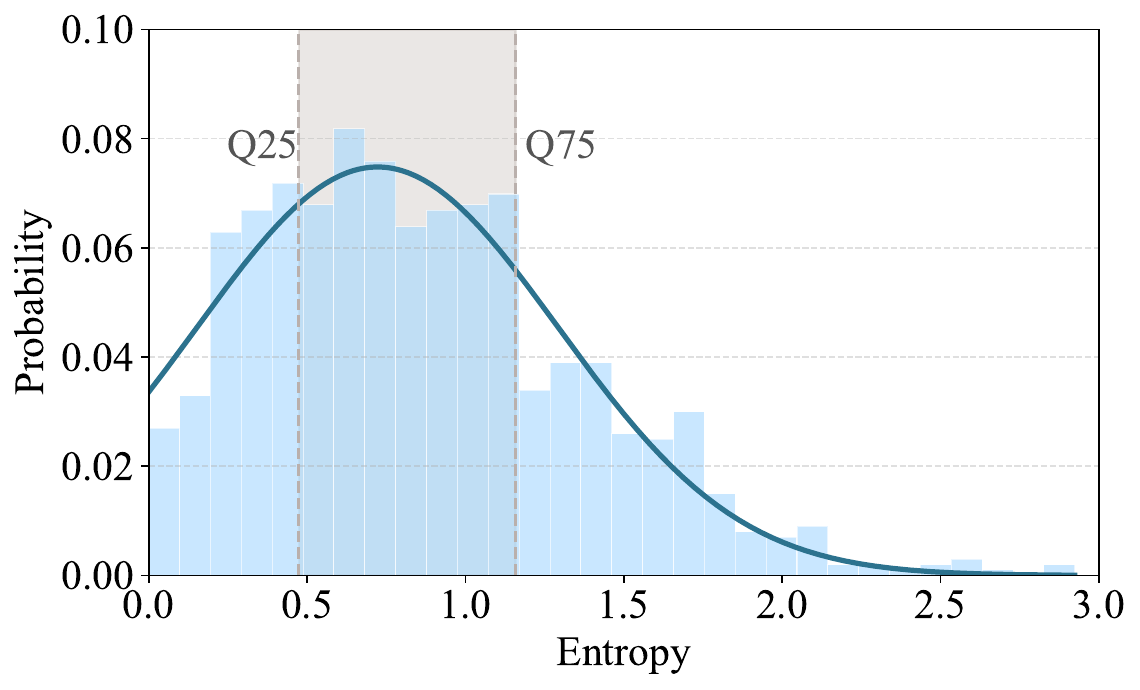}
  \caption{Tool entropy distribution along agent trajectories in SWE-Bench cases}
  \label{fig:tool_entropy}
\end{figure}
We quantify per-step uncertainty in tool selection along an execution trajectory using \emph{tool entropy}:
\begin{equation}
H_{tool}(a_t | s_t) = - \sum_{a \in \mathcal{A}} P(a | s_t) \log P(a | s_t)
\label{eq:1}
\end{equation}
where $P(a \mid s_t)$ denotes the agent policy over a finite tool/action set $\mathcal{A}$ conditioned on the current state $s_t$.

Empirically (Figure~\ref{fig:tool_entropy}), $H_{\text{tool}}$ exhibits a right-skewed distribution, with probability mass concentrated in a low-entropy regime~\cite{wang2024distrl}. This pattern indicates that many trajectory steps correspond to near-deterministic routine operations (e.g., reading or editing files). In contrast, the sparse long tail corresponds to semantically consequential branching points—such as selecting among functionally similar tools—at which uncertainty increases substantially. This distributional imbalance implies that uniform exploration is computationally inefficient. Consequently, an effective TTS paradigm should \textbf{adaptively allocate resources} by reducing computation in high-density, low-entropy regions while prioritizing the infrequent yet high-entropy decision points in the tail.

\subsection{Self-Deceptive Debugging}
\label{sec:self_deceptive}
\begin{figure}[t]
  \centering
  \includegraphics[width=0.5\linewidth]{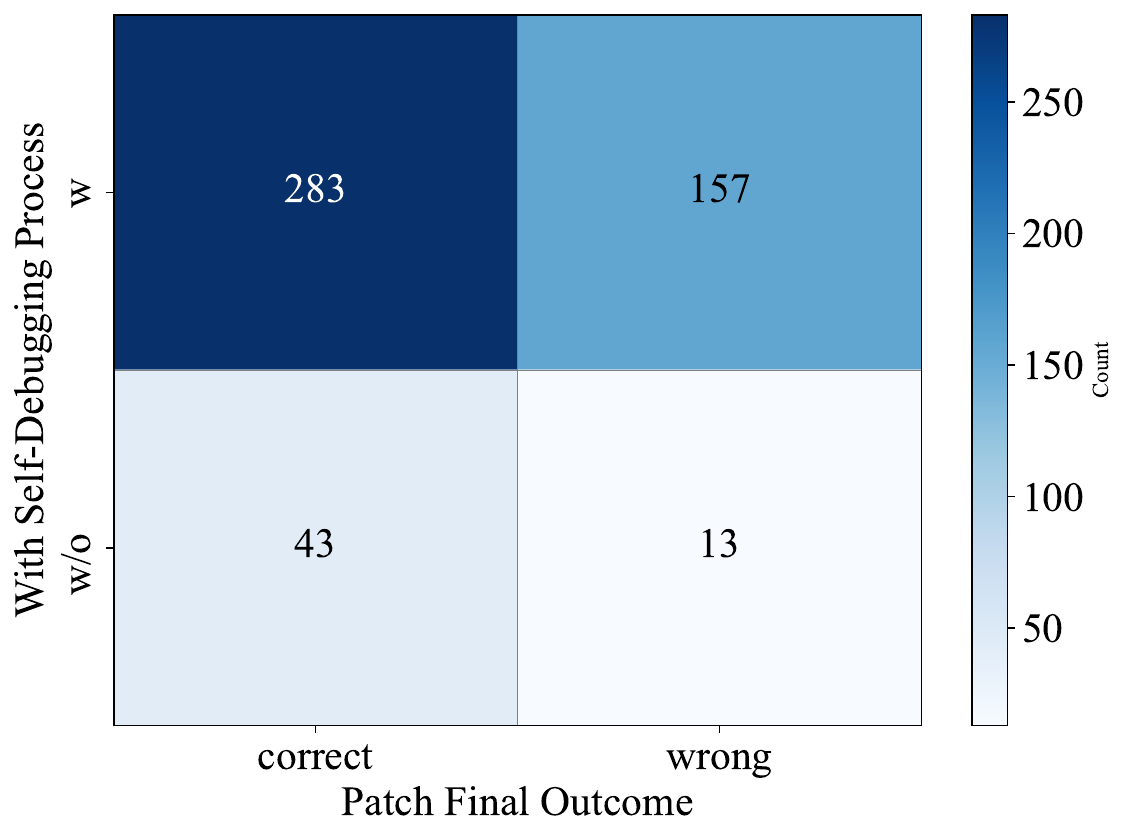}
  \caption{Trajectory-Aware Analysis of Debugging Processes in Autonomous Agents on SWE-Bench-Verified, Using Kimi-K2-Instruct as the base model}
  \label{fig:deceptive_debugging}
\end{figure}
We further investigate behavioral patterns during task execution. As shown in Figure \ref{fig:deceptive_debugging}, on the SWE-Bench-Verified benchmark~\cite{jimenez2023swe} comprising 500 tasks, 88.0\% (440/500) of trajectories exhibit explicit self-verification procedures, wherein the agent proactively generates tests, executes them, and iteratively refines the proposed patch based on the resulting feedback. Notably, among these 440 trajectories that incorporate self-verification, 35.7\% (157/440) nevertheless fail to yield a correct patch. This finding indicates that existing single-trajectory self-verification is frequently confined to a single reasoning perspective and lacks multidimensional assessment of patch correctness and robustness.

We refer to this failure mode as \emph{self-deceptive debugging}: the agent nominally executes a verification routine, yet substantively accepts an invalid patch due to cognitive limitations or entrapment in local optima, ultimately resulting in unsuccessful repairs. The prevalence of self-deceptive debugging highlights a fundamental limitation of single-trajectory self-correction: verification along a single execution path is insufficient to expose latent defects. Because code correctness is inherently multi-dimensional (e.g., functional completeness, boundary robustness, and behavioral consistency), single-view verification is susceptible to confirmation bias. This observation motivates a multi-view, cross-trajectory verification paradigm, in which multiple candidate patches are generated and cross-validated to improve reliability.
\subsection{Problem Formulation}
We formalize the core problem as follows: given a software engineering task, design an agent architecture that (1) dynamically allocates computation based on per-step decision uncertainty—thereby avoiding redundant exploration in near-deterministic steps—and (2) leverages debugging signals from multiple trajectories to synthesize a comprehensive, multi-dimensional test suite for cross-validation, thereby reducing self-deceptive debugging and improving patch quality. These principles form the foundation of {Entropy-guided Stepwise Scaling}, detailed in Section~\ref{sec:method}.

\section{Method}
\label{sec:method}
\begin{figure*}[t]
  \centering
  \includegraphics[width=\textwidth]{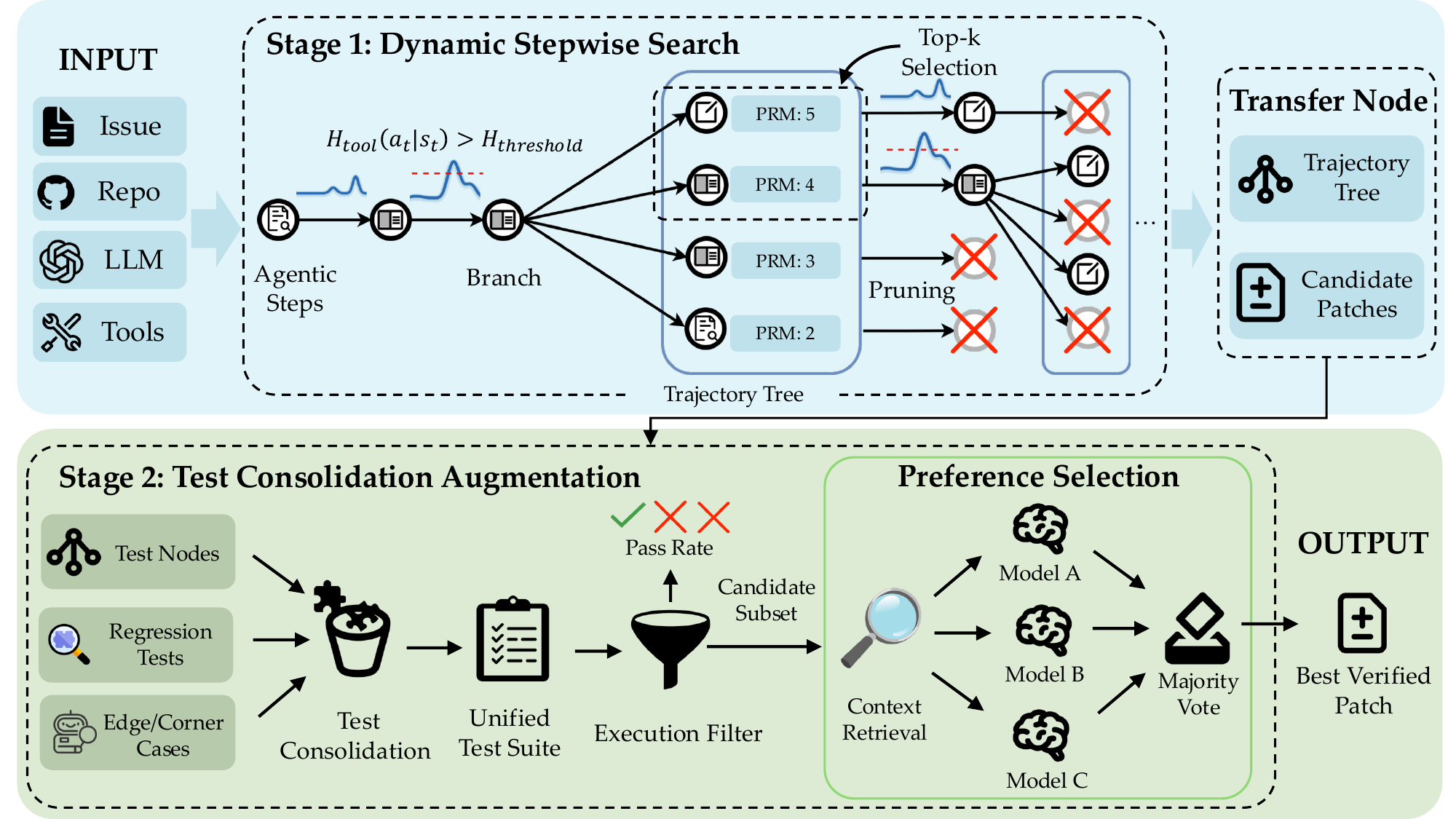}
  \caption{Overview of Entropy-guided Stepwise Scaling}
  \label{fig:tts_overview}
\end{figure*}

\subsection{Overview}
As illustrated in Figure~\ref{fig:tts_overview}, we propose EGSS, a framework that generates high-quality and robust code patches under a constrained inference budget.
Specifically, in the first stage, the agent dynamically identifies key decision points during inference based on tool entropy~\cite{dong2025agentic} and conducts targeted exploration at these critical nodes, thereby substantially increasing the diversity and semantic coverage of candidate trajectories while preserving generation quality.
In the second stage, by leveraging debugging signals embedded across multiple trajectories, the framework constructs an executable, consolidated test suite, and integrates context-aware preference selection with an ensemble voting mechanism across multiple models to systematically validate candidate patches and filter them for robustness.
This two-stage design constitutes a closed-loop optimization process: the first stage provides a high-potential candidate set, whereas the second stage offers a reliable selection mechanism over multiple candidate sets.

\subsection{Dynamic Stepwise Search}
Motivated by the observation that uncertainty is not uniformly distributed across decision points in tool-augmented reasoning, we aim to prioritize computation at steps that are most ambiguous. Specifically, we identify critical steps via Eq.~\ref{eq:1}. When the entropy is high—indicating substantial uncertainty over the appropriate tool choice—we invoke an auxiliary LLM-based judge~\cite{li2025generationjudgmentopportunitieschallenges} to evaluate candidate continuations and prune low-quality branches. This adaptive mechanism allocates inference budget more effectively than repeat sampling strategies~\cite{antoniades2024swe}. Overall, this \emph{stepwise focusing with dynamic evaluation} yields a favorable performance--cost trade-off.

In Dynamic Stepwise Search, the judge provides trajectory-level feedback that is incorporated into the search objective to select the top-$K$ partial paths~\cite{zhang2024rest}. For a partial trajectory $\tau_t = (s_0, a_0, \dots, s_{t-1}, a_{t-1})$, we define
\begin{equation}
S(\tau_t) = \frac{1}{\text{length}(\tau_t)^\gamma} \left[ \sum_{i=0}^{t-1} \left( \log P(a_i \mid s_i) \right. \right. \left. \left. \vphantom{\sum_{i=0}^{t-1}} + \lambda \cdot J(a_i \mid \tau_{i-1}, s_i) \right) \right]
\end{equation}
where $s_i$ denotes the step-$i$ state (encoding the available history/context) and $a_i$ is the
corresponding action. $\log P(a_i\mid s_i)$ is the base model log-likelihood, while
$J(\cdot)$ is the judge-provided evaluation signal; $\lambda\in[0,1]$ controls their relative
contribution (we use $\lambda\in[0.6,1.0]$ in practice). The length penalty
$\mathrm{length}(\tau_t)^\gamma$ with $\gamma\ge 0$ normalizes for trajectory length to mitigate
bias toward short trajectories; $\gamma=0$ recovers the unnormalized sum, and larger $\gamma$
approximates per-step averaging. DSS ranks partial trajectories by $S(\tau_t)$ and prioritizes
expansion of those actions $a_i$ with higher estimated utility, thereby focusing computation on informative
decision points and reducing wasted tokens on low-value or repetitive continuations.

To obtain a reliable and informative $\text{JudgeScore}$, we employ a trained LLM as a judge~\cite{li2025generationjudgmentopportunitieschallenges} and perform online evaluation after each expansion step. The judge input is:
\begin{equation}
I_{\text{Judge}}=\{\text{Task},\,\tau_{\le t},\,\text{Rubric},\,a_i\}
\end{equation}
where $\text{Task}$ denotes the problem specification, $\tau_{\le t}$ the current partial trajectory (i.e., the action/tool sequence up to step $t$), and $\text{Rubric}$~\cite{kimiteam2025kimik2openagentic} the explicit evaluation criteria (see Appendix~\ref{app:dss} for details). The judge score at step-$i$ can be computed as
\begin{equation}
J (a_i \mid \tau_{t-1}, s_i) = \mathcal{J}_{\text{LLM}}(I_{\text{Judge}})
\end{equation}

\subsection{Test Consolidation Augmentation}
\begin{algorithm}[t]
\small
\caption{Test Consolidation Augmentation}
\label{alg:tca}
\begin{algorithmic}[1]
\Require Trajectory tree $\mathcal{T}$ from DSS, Candidate patches $\mathcal{P}$, Codebase $\mathcal{R}$, pass rate threshold $\tau$
\Ensure Selected patch $p^*$

\State \textbf{// Step 1: Extract debugging actions}
\State $\mathcal{A}_{\text{debug}} \gets \emptyset$
\For{each node $v$ in $\mathcal{T}$}
    \If{$v$ invokes debugging tool}
        \State $\mathcal{A}_{\text{debug}} \gets \mathcal{A}_{\text{debug}} \cup \{v\}$
    \EndIf
\EndFor

\State \textbf{// Step 2: Generate consolidated test suite}
\State $\mathcal{S} \gets \textsc{TestConsolidationAgent}(\mathcal{A}_{\text{debug}}, \mathcal{R})$

\State \textbf{// Step 3: Evaluate patches with test suite}
\State $\mathcal{P}_{\text{valid}} \gets \emptyset$
\For{each patch $p$ in $\mathcal{P}$}
    \State $r \gets \textsc{TestEvaluationAgent}(p, \mathcal{S}, \mathcal{R})$
    \If{$r.\texttt{passRate} \geq \tau$}
        \State $\mathcal{P}_{\text{valid}} \gets \mathcal{P}_{\text{valid}} \cup \{p\}$
    \EndIf
\EndFor

\State \textbf{// Step 4: Select best patch via voting}
\State $\mathcal{V} \gets \emptyset$
\For{each model $m$ in base models}
    \State $p_m \gets \textsc{PreferenceSelectorAgent}_m(\mathcal{P}_{\text{valid}}, \mathcal{R})$
    \State $\mathcal{V} \gets \mathcal{V} \cup \{p_m\}$
\EndFor
\State $p^* \gets \textsc{MajorityVote}(\mathcal{V})$

\State \Return $p^*$
\end{algorithmic}
\end{algorithm}

The phenomenon of {self-deceptive debugging}, wherein agents perform verification procedures yet still accept incorrect patches (Section~\ref{sec:self_deceptive}), exposes a fundamental limitation of conventional Test-Time Scaling (TTS) pipelines: their dependence on a narrow, unidimensional correctness criterion. By formulating patch validation as an isolated terminal decision, these pipelines fail to exploit the diagnostic evidence distributed across diverse debugging trajectories. This restricted perspective not only degrades the robustness of patch selection but also neglects the collective information afforded by multi-trajectory exploration, which could otherwise reveal subtle defects that remain undetected under superficial test success.

To directly address this limitation, we propose \textbf{Test Consolidation Augmentation (TCA)}, a robust and reliable patch selection pipeline that consolidates fragmented debugging evidence into a unified and executable verification standard. Unlike LLM-as-Judge heuristics~\cite{li2025generationjudgmentopportunitieschallenges}, which are susceptible to inconsistency and hallucination, TCA anchors final decisions in concrete and reproducible test outcomes synthesized from the complete set of trajectories.

As outlined in Algorithm~\ref{alg:tca}, TCA operates in four consecutive stages:

\paragraph{Debugging action extraction.}
Given the trajectory tree $\mathcal{T}$ generated by \textbf{DSS}, we first collect all nodes that invoke debugging tools (e.g., test generation/execution). These nodes constitute $\mathcal{A}_{\text{debug}}$, which compactly summarizes the debug-related tool-invocation steps across trajectories.

\paragraph{Consolidated test suite generation.}
We then employ \textsc{TestConsolidationAgent}, an agent equipped with the same MCP tools \cite{abdelaziz-etal-2024-granite} used in the code generation(see Appendix~\ref{app:tca} for details), to synthesize an executable test suite $\mathcal{S}$ from $\mathcal{A}_{\text{debug}}$ and the repository $\mathcal{R}$. Concretely, the agent consolidates complementary testing intents encoded in $\mathcal{A}_{\text{debug}}$ and grounds them in the repository context, yielding a unified suite that provides broader and more robust coverage than any single-trajectory test set.

\paragraph{Patch evaluation and filtering.}
Next, we apply \textsc{TestEvaluationAgent}, configured exclusively with bash MCP tools (see Appendix~\ref{app:tca} for details), to execute $\mathcal{S}$ on each candidate patch $p \in \mathcal{P}$ and compute its pass rate. We retain patches whose pass rate exceeds a threshold $\tau$, producing a high-confidence candidates $\mathcal{P}_{\text{valid}}$. This stage filters out patches that may appear plausible but do not withstand consolidated verification.

\paragraph{Preference selection via voting.}
If multiple candidates pass the consolidated tests, we employ multiple \textsc{Preference Selectors} to vote for a single best patch. Specifically, each selector independently selects one patch from $\mathcal{P}_{\text{valid}}$ conditioned on the repository context $\mathcal{R}$.

\section{Experiments}
\subsection{Benchmark and Evaluation Metrics}
We evaluate the effectiveness of {Entropy-guided Stepwise Search} on the SWE-Bench benchmark~\cite{jimenez2023swe}, adopting the \textbf{resolved percentage}---defined as the proportion of benchmark instances for which a correct patch is successfully generated---as the primary metric for quantitatively comparing EGSS with established baselines. In addition, we report the \textbf{Oracle} metric~\cite{zhang2024diversity}, which considers an instance resolved if at least one correct patch is identified among the \(N\) generated candidate patches. Together, these metrics offer a comprehensive evaluation of both the empirical effectiveness of EGSS and its upper-bound performance under candidate selection.

\subsection{Compared TTS Method}
To rigorously evaluate the effectiveness of the proposed method, we conduct a comparative study against several widely used TTS baselines, with emphasis on two critical stages: \textit{sampling} and \textit{patch selection}.

In the sampling stage, we investigate \textbf{repeat sampling}~\cite{gao2025trae} under controlled stochasticity by adjusting generation hyperparameters (e.g., temperature), thereby encouraging diversity in the synthesized code patches and the corresponding execution traces.

For patch selection, we adopt two widely used strategies—\textbf{Deibase}~\cite{zhang2024diversity} and \textbf{Augment}~\cite{chen2025swebench}—as baseline methods for comparison. Following their original formulations, we further refine and extend both approaches:

\begin{itemize}
    \item \textbf{Deibase}~\cite{zhang2024diversity}: The original method retrieves relevant contextual information from the codebase and, together with each candidate patch and the issue description, prompts an LLM-as-judge~\cite{li2025generationjudgmentopportunitieschallenges} to assign a score to each candidate. We extend this design by adopting an Agent-as-Judge~\cite{zhuge2024agentasajudgeevaluateagentsagents} paradigm, in which a judge agent autonomously evaluates each candidate patch by constructing targeted test suites and/or performing rubric-guided code reviews.
    
    \item \textbf{Augment}\cite{chen2025swebench}: The original method employs an LLM-as-judge paradigm~\cite{li2025generationjudgmentopportunitieschallenges}, prompting the model to assess each candidate patch against the issue description and select the most appropriate one. We adapt this framework to our selector design by aggregating the decisions of multiple selectors via majority voting, thereby producing a single consensus patch.
\end{itemize}

\subsection{Experiment Setup}
In our experiments, the base agent framework integrates a suite of MCP tools~\cite{abdelaziz-etal-2024-granite} (see Appendix~\ref{app:code_gen} for agent details). For patch generation, we evaluate several large language models (LLMs): Kimi-K2-Instruct~\cite{kimiteam2025kimik2openagentic}, GLM-4.6, and GLM-4.5-Air~\cite{5team2025glm45agenticreasoningcoding}.

During the Dynamic Stepwise Search stage, all models in our experiments operate with a temperature of 1, and the stepwise rollout count is set to 4. Specifically, at high-entropy steps, we resample candidate tool-call outcomes four times. A fine-tuned Qwen3-8B~\cite{yang2025qwen3technicalreport} model serves as the judge, providing scalar scoring signals.  We further evaluate sensitivity to the ensemble size by comparing configurations with 4, 6, and 8 rollouts, which correspond to generating 4, 6, and 8 final patch candidates, respectively.

In the Test Consolidation Augmentation stage, Kimi-K2-Instruct~\cite{kimiteam2025kimik2openagentic} serves as the base model for both the \textsc{TestConsolidationAgent} and the \textsc{TestEvaluationAgent}. Final decisions are determined via majority voting over an ensemble of expert models comprising Kimi-K2-Instruct~\cite{kimiteam2025kimik2openagentic}, GLM-4.6~\cite{5team2025glm45agenticreasoningcoding}, and Qwen3-Coder-480B-A35B-Instruct~\cite{yang2025qwen3technicalreport}.

\subsection{Experimental Results}
\begin{table}[t]
\centering
\small
\setlength{\tabcolsep}{4pt}
\begin{tabular}{l l cc}
\toprule
\multirow{2}{*}{\textbf{Method}} & \multirow{2}{*}{\textbf{Base Model}} & \multicolumn{2}{c}{\textbf{Resolved\%}} \\
\cmidrule(lr){3-4}
& & \begin{tabular}[c]{@{}c@{}}Swe-Bench\\Verified\end{tabular} & \begin{tabular}[c]{@{}c@{}}Swe-Bench\\Lite\end{tabular} \\
\midrule
\multirow{5}{*}{Single Shot} & Kimi-K2-Instruct & 63.2\% & 48.33\% \\
 & GLM-4.6 & 65.8\% & 51\% \\
 & GLM-4.5-Air & 56.4\% & N/A \\
 & Claude-4-sonnet & 68\% & 57\% \\
 & Claude-45-sonnet & 70.4\% & 62\% \\
\midrule
\multirow{3}{*}{Repeat+Dei Aug} & Kimi-K2-Instruct & 65.4\% & N/A \\
 & GLM-4.6 & 66.6\% & N/A \\
 & GLM-4.5-Air & 57.2\% & N/A \\
\midrule
\multirow{3}{*}{\textbf{EGSS (Ours)}} & Kimi-K2-Instruct & 70.6\% & 61\% \\
 & GLM-4.6 & \textbf{73.8\%} & \textbf{64\%} \\
 & GLM-4.5-Air & \textbf{62.6\%} & N/A \\
\bottomrule
\end{tabular}
\caption{Performance differences across models under various methods, with fixed Ensemble size \( K = 4 \)}
\label{tab:main_results}
\end{table}
As summarized in Table~\ref{tab:main_results}, with an ensemble size of \(K=4\), {EGSS} achieves 73.8\% on SWE-Bench-Verified and 64\% on SWE-Bench-Lite. Across different base LLMs, EGSS consistently outperforms the compared TTS-based methods, delivering relative improvements of approximately 5\%--10\%.

To further elucidate the sources of these gains, we structure our ablation study into two stages corresponding to EGSS’s pipeline. The first stage investigates the sampling component, comparing Dynamic Stepwise Search (DSS) with conventional repeat sampling~\cite{gao2025trae} in terms of token efficiency and oracle performance~\cite{zhang2024diversity}. The second stage assesses the effect of Test Consolidation Augmentation under fixed ensemble sizes across sampling strategies,  examining its impact on the stability of patch generation. Detailed analyses are presented in the following sections.

\subsubsection{Analysis of Token Usage}
\begin{figure}[t]
  \centering
  \includegraphics[width=\linewidth]{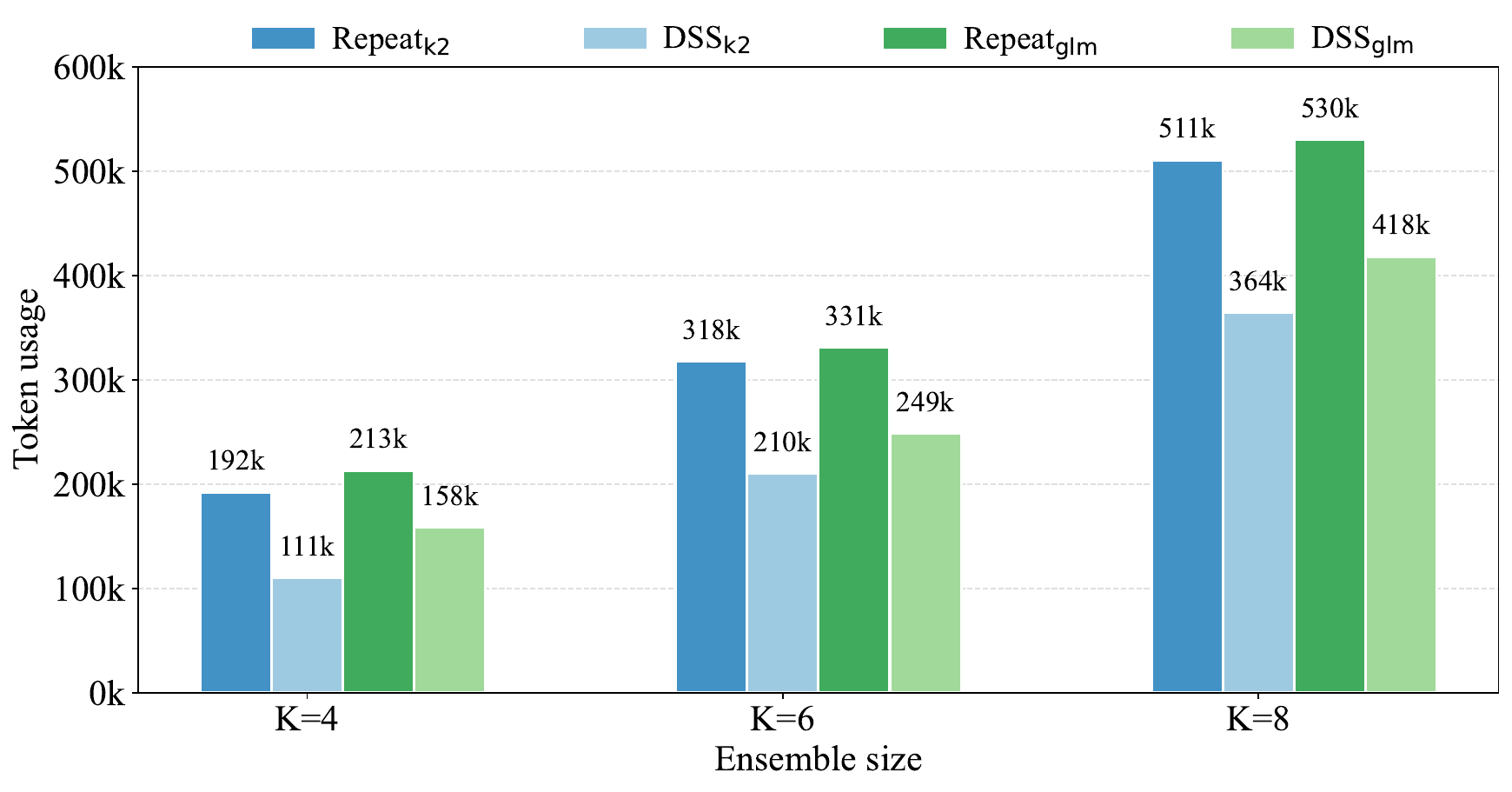}
  \caption{Average token usage per instance on the SWE-Bench benchmark, aggregated across different ensemble sizes and sampling strategies for Kimi-K2-Instruct and GLM-4.6.}
  \label{fig:token_usage}
\end{figure}

\begin{table}[htbp]
\small
\centering
\label{tab:dss_results}
\begin{tabular}{@{}llll@{}}
\toprule
Base Model & K & Sampling Strategy & Oracle \\ \midrule
\multirow{6}{*}{Kimi-K2-Instruct} & \multirow{2}{*}{4} & Repeat & 71.8\%\\
 &  & DSS & 73.8\%\\ \cmidrule(l){2-4} 
 & \multirow{2}{*}{6} & Repeat & 73.4\% \\
 &  & DSS & 75.6\%\\ \cmidrule(l){2-4} 
 & \multirow{2}{*}{8} & Repeat & 74.2\% \\
 &  & DSS & 78.8\% \\ \midrule
\multirow{6}{*}{GLM--4.6} & \multirow{2}{*}{4} & Repeat & 73.4\%\\
 &  & DSS & 77.4\%\\ \cmidrule(l){2-4} 
 & \multirow{2}{*}{6} & Repeat & 74.8\% \\
 &  & DSS & 78.4\%\\ \cmidrule(l){2-4} 
 & \multirow{2}{*}{8} & Repeat & 76.6\% \\
 &  & DSS & 79.6\%\\ \midrule
\multirow{2}{*}{GLM--4.5--Air} & \multirow{2}{*}{4} & Repeat & 60.8\%\\
 &  & DSS & 64.8\% \\ \bottomrule
\end{tabular}
\caption{Comparison of different sampling strategies on SWE-Bench-Verified}
\label{tab:sample_method_compared}
\end{table}

\begin{table*}[t]
\centering
\small 
\begin{tabular}{llcccc}
\toprule
\textbf{Base Model} & \textbf{K} & \multicolumn{4}{c}{\textbf{Resolved \%}} \\
\cmidrule(lr){3-6}
& & \textbf{Repeat + Dei Aug} & \textbf{Repeat + TCA} & \textbf{DSS + Dei Aug} & \textbf{DSS + TCA (EGSS)} \\
\midrule
\multirow{3}{*}{Kimi-K2-Instruct} & 4 & 65.4\% & 66.6\% & 65.8\% & 68.4\% \\
 & 6 & 66\% & 67.6\% & 68.2\% & 70.6\% \\
 & 8 & 66.4\% & 67.6\% & 70.0\% & \textbf{72.2\%} \\
\midrule
\multirow{3}{*}{GLM--4.6} & 4 & 64.2\% & 66.4\% & 71.6\% & 73.8\% \\
 & 6 & 67\% & 68.6\% & 70.2\% & 72\% \\
 & 8 & 68.2\% & 70.4\% & 71.0\% & \textbf{74.6\%} \\
\bottomrule
\end{tabular}
\caption{Performance comparison (Resolved \%) across different TTS methods, base models, and Ensemble size \(K\) values on SWE-Bench-Verified}
\label{tab:tca-comparison}
\end{table*}

\begin{figure}[t]
  \centering
  \begin{subfigure}[b]{0.48\linewidth}
    \centering
    \includegraphics[width=\linewidth]{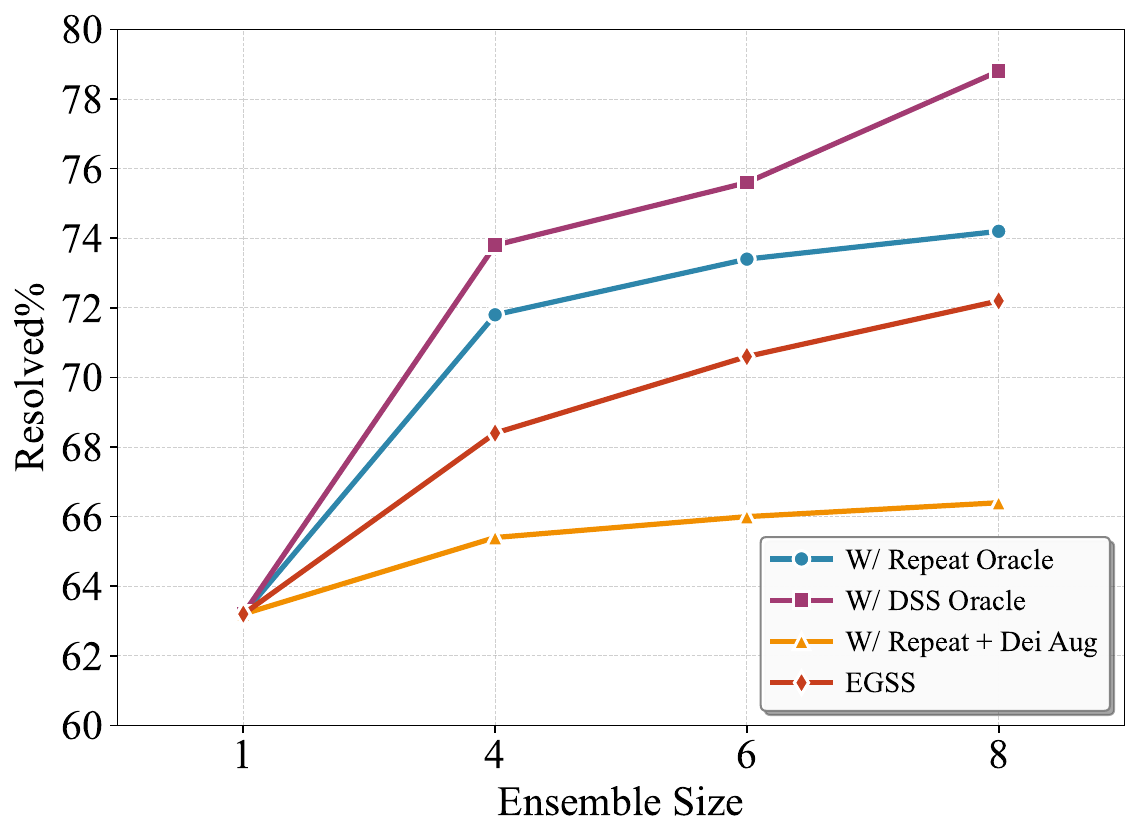}
    \caption{Kimi-K2-Instruct}
    \label{fig:k2_rollout_ana}
  \end{subfigure}
  \hfill
  \begin{subfigure}[b]{0.48\linewidth}
    \centering
    \includegraphics[width=\linewidth]{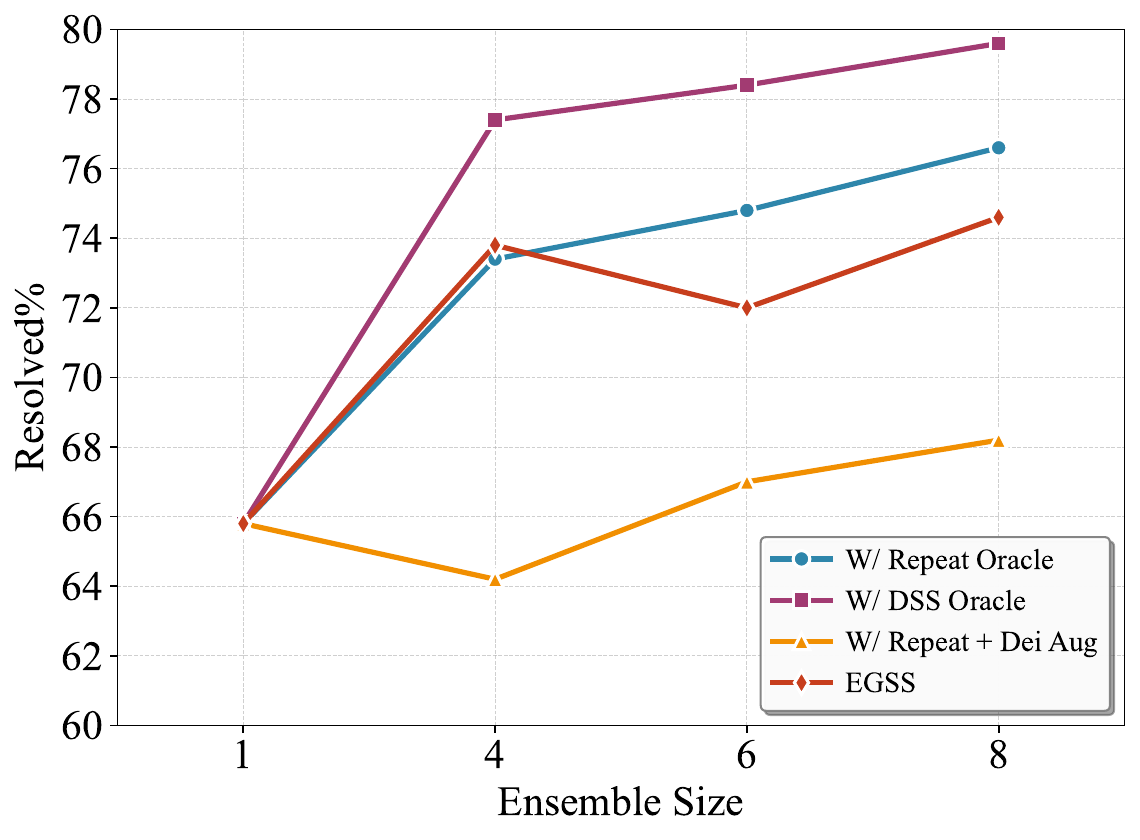}
    \caption{GLM-4.6}
    \label{fig:rollout_analysis}
  \end{subfigure}
  \caption{Comparison of ensemble strategies across different ensemble sizes}
  \label{fig:glm_rollout_ana}
\end{figure}

As illustrated in Figure~\ref{fig:token_usage} and Table~\ref{tab:sample_method_compared}, DSS yields an average \textbf{28\%} reduction in stable token consumption, averaged over the evaluated ensemble sizes (\(K = 4, 6, 8\)), relative to the conventional repeat sampling strategy. A more fine-grained analysis further suggests that this reduction is not merely attributable to cost savings, but also reveals an inherent scalability limitation of repetition-based approaches. Specifically, as \(K\) increases, token usage under repeat sampling increases disproportionately and sharply, indicating a substantial deterioration in computational efficiency for complex tasks that require larger search spaces. In contrast, DSS alleviates this cost escalation via a principled and adaptive sampling mechanism.

Importantly, DSS not only reduces token consumption but also attains superior task performance—even, even under smaller ensemble sizes. In particular, DSS with \(K = 4\) achieves performance comparable to repeat sampling with \(K = 8\), indicating that DSS provides markedly higher sample efficiency without sacrificing solution quality. Collectively, these results support the conclusion that DSS substantially improves the pass@\(k\) metric while simultaneously reducing the required ensemble size.

\subsubsection{Analysis of Patch Selection}

As shown in Table~\ref{tab:tca-comparison} and Figure~\ref{fig:rollout_analysis}, Test Consolidation Augmentation demonstrates consistently strong performance across ensemble sizes \(K\) and sampling strategies. In particular, TCA exhibits both efficiency and robustness: using GLM-4.6 under the DSS strategy, it attains a success rate of \(73.8\%\) at \(K=4\), which already exceeds the strongest baseline (DSS+Dei Aug), and further improves to \(74.6\%\) at \(K=8\).

Unlike methods for which increasing \(K\) can introduce additional noise and thereby impair patch selection—reflected in non-monotonic performance trends—TCA effectively exploits larger ensembles. It approaches the Oracle upper bound with fewer samples while preserving a higher-quality candidate pool, thereby supporting more reliable patch selection as \(K\) increases.

\section{Conclusion}

In this paper, we propose Entropy-guided Stepwise Scaling (EGSS), a novel TTS framework designed to enhance both computational efficiency and operational reliability in autonomous software engineering. EGSS dynamically allocates computational resources to high-uncertainty steps based on tool entropy~\cite{dong2025agentic}, and integrates cross-trajectory test evidence to enable robust patch selection. Empirically, EGSS attains state-of-the-art performance on SWE-Bench while reducing token consumption by more than 28\%. These results indicate that a principled stepwise scaling strategy can substantially improve the practical applicability of large language models (LLMs) to real-world software engineering tasks, without increasing the ensemble size.

\appendix
\clearpage

\begingroup 
\raggedbottom 
\setlength{\parskip}{4pt} 
\section{Dynamic Stepwise Search}
Implementation Detail of Dynamic Stepwise Search
\label{app:dss}
\subsection{Code Generation}
\label{app:code_gen}
First we will introduce our code agent used for code generation, the system prompt used are as following, and Tools available for Agent are:
\begin{itemize}
    \item \texttt{read\_file}: Read file contents with optional line range selection
    \item \texttt{write\_file}: Create or overwrite files
    \item \texttt{edit\_file}: Perform edits via search-and-replace
    \item \texttt{grep}: Fast code search powered by ripgrep
    \item \texttt{glob}: File discovery using glob patterns
    \item \texttt{bash}: Execute shell commands with timeout control
\end{itemize}
\begin{inputbox}{System Prompt}
\small
\begin{verbatim}
${available_tools}
\end{verbatim}
You are an AI coding assistant designed to help developers with their coding tasks. You have access to tools that allow you to read and write files in the workspace.

Your approach: \\
1. Carefully analyze the user's request \\
2. Use available tools to gather necessary information \\
3. Propose clear, well-thought-out solutions \\
4. Execute changes carefully and verify results \\

When modifying files: \\
- Always read files before modifying them \\
- Make precise, targeted changes \\
- Explain what you're doing and why \\

Be concise, accurate, and helpful.
\end{inputbox}

\subsection{Judge Model}
When an Agent step is identified as a high-entropy action, we roll out multiple actions in the current step. To determine the preference order among these actions, we introduce a Judge Model. The Judge Model takes the current task, a partial trajectory of the agent, and a rubric as input. The detailed system prompt and the rubric used in our experiments are provided below.
\begin{inputbox}{System Prompt}
\small
\textbf{Role}
Please act as an impartial judge to \textbf{verify} the patch provided by assistant to the user's task below.

\textbf{Instruction}
\begin{itemize}
    \item User will provide you with a response and the corresponding task, and your task is to verify its rationality with the rubric given below.
    \item Base your decision solely on how well the response addresses the user’s question and adheres to the assistant instructions.
    \item You can use the available tools to assess the provided rubric
    \item \textbf{Important}, If there are anything wrong with the provided response, \textbf{do not} try to fix it!
\end{itemize}

\textbf{Rubric}
Your evaluation should focus on the following criteria:
\begin{verbatim}
${rubric}
\end{verbatim}

\textbf{Output format}
\textbf{IMPORTANT}
Your final reply must be structured in the Json format consist of one args
\begin{itemize}
    \item critique, required, str
    \item scalar, required, str, this is a fractional string where the numerator represents the score and the denominator is the total possible score from the rubric.
\end{itemize}

Here is an example of the output
\begin{verbatim}
{
    "critique": "xxxx",
    "scalar": "a/b"
}
\end{verbatim}

\textbf{User Task}:
\begin{verbatim}
${task}
\end{verbatim}

\textbf{Agent History Trajectory}
\begin{verbatim}
${trajectory}
\end{verbatim}

\textbf{Current Round Response}
\begin{verbatim}
${response}
\end{verbatim}

\textbf{Output}
Now it's your turn.
\end{inputbox}

A rubric is an evaluation criterion. In our experiment, we focused the rubric primarily on the reasonableness of certain processes, such as expected tool usage.

\begin{inputbox}{Rubric}
\small 
\textbf{1. Step Consistency}
\textbf{Definition:} Assesses whether the entire execution trace is logically coherent, free of contradictions, and devoid of redundancy.

\rubrictable{
    \textbf{0} & \textbf{Inconsistent} \\
    \textbf{1} & \textbf{Basically Consistent} \\
    \textbf{2} & \textbf{Partially Consistent} \\
    \textbf{3} & \textbf{Fully Consistent}
}

\textbf{2. Context Awareness}
\textbf{Definition:} Evaluates whether the agent effectively utilizes and dynamically updates historical context, avoiding repetition or conflict.

\rubrictable{
    \textbf{0} & \textbf{Context Ignored} \\
    \textbf{1} & \textbf{Partially Contextualized} \\
    \textbf{2} & \textbf{Mostly Contextualized} \\
    \textbf{3} & \textbf{Fully Contextualized}
}

\textbf{3. Goal Prioritization}
\textbf{Definition:} Determines whether the agent correctly identifies and prioritizes critical sub-goals, preventing resource misallocation.

\rubrictable{
    \textbf{0} & \textbf{Chaotic Prioritization} \\
    \textbf{1} & \textbf{Unbalanced Prioritization} \\
    \textbf{2} & \textbf{Adequate Prioritization} \\
    \textbf{3} & \textbf{Optimal Prioritization}
}

\textbf{4. Expected Tool Use}
\textbf{Definition:} Selection, invocation, and sequencing of the \textit{correct} tools with \textit{appropriate} arguments and \textit{minimal} redundancy, including proper error handling and validation. This includes ensuring format-specific features are fully implemented across all relevant code paths, and critically, handling \textbf{edge cases} and \textbf{optimization paths} in the target codebase (e.g., fast-delete optimization in Django deletion logic).

\rubrictable{
    \textbf{0} & \textbf{Misuse or Abstain} \\
    \textbf{1} & \textbf{Partial Tool Use} \\
    \textbf{2} & \textbf{Adequate Tool Use} \\
    \textbf{3} & \textbf{Optimal Tool Use}
}

\textbf{5. Diagnostic Precision}
\textbf{Definition:} Evaluates how effectively the agent isolates the root cause of an issue, distinguishes between symptoms and causes, and provides targeted explanations or fixes without over-generalizing or misidentifying the problem.

\rubrictable{
    \textbf{0} & \textbf{Misdiagnosis} \\
    \textbf{1} & \textbf{Partial Diagnosis} \\
    \textbf{2} & \textbf{Accurate Diagnosis} \\
    \textbf{3} & \textbf{Precise Diagnosis}
}
\end{inputbox}

A case study of the Judge Model is presented below, illustrating how preference ordering among high-entropy actions is determined. Due to space constraints, only a portion of the input is displayed here, with certain content omitted and indicated by ellipses.
\begin{casebox}{Case Input: \$\{task\}}
\small
Modeling's separability\_matrix does not compute separability correctly for nested Compound Models \\ 
Consider the following model:
...
\end{casebox}
\begin{casebox}{Case Input: \$\{trajectory\}}
\small
\begin{lstlisting}[
    basicstyle=\ttfamily\small,
    columns=flexible,
    breaklines=true,
    tabsize=1, % 关键：设置缩进宽度为2个空格
    showstringspaces=false
]
<trajectory>
    <tool_call name="read_file">
        <arguments>
            {"path": "/testbed/astropy/modeling/tests/test_separable.py"}
        </arguments>
        <response>
            ...
        </response>
    </tool_call>
    <tool_call name="bash">
		      <arguments>
            {"command": "cd /testbed && python -m pytest astropy/modeling/tests/test_separable.py -v"}
        </arguments>
        <response>
            Command executed successfully.\n\nOutput:...
        </response>
	   </tool_call>
    ...
</trajectory>
\end{lstlisting}
\end{casebox}

\begin{casebox}{Case Input: \$\{response\}}
\small
\begin{lstlisting}[
    basicstyle=\ttfamily\small,
    columns=flexible,
    breaklines=true,
    tabsize=1, % 关键：设置缩进宽度为2个空格
    showstringspaces=false
]
<response>
    <tool_call name="edit_file">
        <arguments>
            {
                "file_path": "/testbed/astropy/modeling/tests/test_separable.py", 
                "old_string": ..., 
                "new_string": ...
            }
        </arguments>
    </tool_call>
</response>
\end{lstlisting}
\end{casebox}

\begin{outputbox}{Case Judge Model Output}
\small
\begin{lstlisting}
{
    "critique": "The assistant correctly identifies that a new test case is needed to demonstrate the bug reported by the user. The investigative steps, particularly running a debug script to trace the matrix creation, were relevant. However, the overall process was inefficient and illogical....",
    "scalar": "18/30"
}
\end{lstlisting}
\end{outputbox}

\section{Test Consolidation Augmentation}
\label{app:tca}

This section will reveal the implementation details of Test Consolidation Augmentation

\subsection{Test Consolidate Agent}
The Test Consolidation Agent is designed to consolidate debugging signals from diverse agent trajectories. It shares the same MCP tools as the Code Generation agent (introduced in Section~\ref{app:code_gen}) and possesses equivalent privileges, granting it full access to any file within the repository.

Tools available for Test Consolidate Agent are:
\begin{itemize}
    \item \texttt{read\_file}: Read file contents with optional line range selection
    \item \texttt{write\_file}: Create or overwrite files
    \item \texttt{edit\_file}: Perform edits via search-and-replace
    \item \texttt{grep}: Fast code search powered by ripgrep
    \item \texttt{glob}: File discovery using glob patterns
    \item \texttt{bash}: Execute shell commands with timeout control
\end{itemize}

\begin{inputbox}{System Prompt}
\small

\begin{verbatim}
${available_tools}
\end{verbatim}
\textbf{ROLE:}
Act as an expert of programming tester. Given a codebase, an issue, and trajectories generated by different code agents, you need to generate \textbf{a single test file} that can be used to validate the patch generated.

\textbf{WORK PROCESS:}
\begin{enumerate}[nosep, leftmargin=*, label=\arabic*., itemsep=0pt, parsep=0pt] 
    \item \textbf{Contextual Analysis:} Comprehend the issue and relevant codebase components (referenced code, patched code, related files, existing regression tests).
    \item \textbf{Trajectory Analysis:} Evaluate test cases and debug code from each trajectory for sufficiency.
    \item \textbf{Coverage Assessment:} Identify and add missing test scenarios (edge cases, regression) to ensure comprehensive coverage.
    \item \textbf{Test Consolidation:} Integrate all test cases (from trajectories, new designs, existing regressions) into `test\_current\_issue.py` at the root.
    \item \textbf{Compilation Assurance:} Ensure `test\_current\_issue.py` compiles without errors, even if tests fail initially.
\end{enumerate}

\textbf{Requirements for TEST FILE:}
\begin{itemize}[nosep, leftmargin=*, itemsep=0pt, parsep=0pt]
    \item \textbf{MUST} be `test\_current\_issue.py` in the root directory.
    \item Provide \textbf{comprehensive} validation for the issue.
    \item \textbf{IMPORTANT:} Executable with `pytest`, reporting total test cases.
\end{itemize}

\textbf{Task}
\begin{verbatim}
${task}
\end{verbatim}

\textbf{Trajectory}
Trajectories sampled from different code agents with only debugging tools:
\begin{verbatim}
${trajectories}
\end{verbatim}

\end{inputbox}

A case study of the Test Consolidation Agent is presented below, illustrating the structure of execution trajectories and providing a concrete example. Due to space constraints, only a portion of the input is displayed here, with certain content omitted and indicated by ellipses.
\begin{casebox}{Case Input: \$\{task\}}
\small
Modeling's separability\_matrix does not compute separability correctly for nested CompoundModels \\ 
Consider the following model:
...
\end{casebox}

\begin{casebox}{Case Input: \$\{trajectories\}}
\small
\begin{lstlisting}[
    basicstyle=\ttfamily\small,
    columns=flexible,
    breaklines=true,
    tabsize=1, % 关键：设置缩进宽度为2个空格
    showstringspaces=false
]
<trajectory idx='1'>
	<tool_call name="write_file">
		<arguments>
    {
        "path": "/testbed/test_issue.py", 
        "content": "..."
    }
		</arguments>
	</tool_call>
	<tool_call name="write_file">
		<arguments>
    {
        "path": "/testbed/debug_issue.py", 
        "content": "..."
    }
		</arguments>
	</tool_call>
	<tool_call name="bash">
		<arguments>
			...
		</arguments>
	</tool_call>
	...
</trajectory>
<trajectory idx='2'>
	....
</trajectory>
\end{lstlisting}
\end{casebox}

\begin{outputbox}{Case Agent Output}
\small
Perfect! Now let me create a final summary of what was done:

\textbf{Summary}

I have successfully created a comprehensive test file that validates the patch for the separability matrix bug in nested CompoundModels. Here's what was accomplished:
....

\textbf{Test Results}

- \textbf{12 comprehensive test cases} - all passed \\
- \textbf{12 existing tests} - all still pass (no regressions) \\
- \textbf{Edge cases} - all validated \\
- \textbf{Original issue examples} - all now work correctly \\

\textbf{Key Test Cases Validated} \\
...
\end{outputbox}

\subsection{Test Evaluation Agent}
The Test Evaluation Agent is designed to execute a specified test file within a strictly controlled environment. It is restricted exclusively to Bash-related Model Context Protocol (MCP) tools and operates solely on the designated test file. To ensure integrity and security, strict filesystem isolation is enforced, preventing access to any files outside the test directory. Upon invocation, the agent automatically executes the test file and autonomously resolves any environmental dependencies or configuration issues that might impede successful execution.

Tools available for Test Evaluation Agent are:
\begin{itemize}
    \item \texttt{bash}: Execute shell commands with timeout control
\end{itemize}

\begin{inputbox}{System Prompt}
\small
\begin{Verbatim}
${available_tools}
\end{Verbatim}
\textbf{ROLE:}
Act as an expert programming tester. Validate code with test\_current\_issue.py based on codebase, issue, and patch.

\textbf{WORK PROCESS:}
\begin{enumerate}[nosep, leftmargin=*, label=\arabic*.] 
    \item \textbf{Execute Test}: Run python test\_current\_issue.py. Output number of solved test cases.
    \item \textbf{Fix Environment Issue (if needed)}: Correct compilation errors in test\_current\_issue.py if not due to patch. \textbf{NOTICE}: you can't modify any codes!
\end{enumerate}

\textbf{OUTPUT FORMAT:}
\textbf{IMPORTANT}: Reply in JSON.
\begin{itemize}[nosep, leftmargin=*] 
    \item \texttt{passed}: int, passed cases.
    \item \texttt{failed}: int, failed cases.
    \item \texttt{error}: int, error cases.
    \item \texttt{total}: int, total test cases.
\end{itemize}
\begin{Verbatim}
{
    "passed": x, 
    "failed": x, 
    "error": x, 
    "total": x
}
\end{Verbatim}

\textbf{Output}
Now it's your turn.
\end{inputbox}

\subsection{Preference Selector}

The Preference Selector is primarily employed to filter and select a single, final patch from a given set of multiple candidates.  It shares the same MCP tools as the Code Generation agent (introduced in Section~\ref{app:code_gen}) and possesses equivalent privileges, granting it full access to any file within the repository.
Tools available for Preference Selector are:
\begin{itemize}
    \item \texttt{read\_file}: Read file contents with optional line range selection
    \item \texttt{write\_file}: Create or overwrite files
    \item \texttt{edit\_file}: Perform edits via search-and-replace
    \item \texttt{grep}: Fast code search powered by ripgrep
    \item \texttt{glob}: File discovery using glob patterns
    \item \texttt{bash}: Execute shell commands with timeout control
\end{itemize}

\begin{inputbox}{System Prompt}
\small
\begin{verbatim}
${available_tools}
\end{verbatim}

\textbf{ROLE:}
Act as an expert code selector. Given a codebase, an github issue and N candidate patches proposed by your colleagues, your responsibility is to select the correct one to solve the issue.

\textbf{WORK PROCESS:}
\begin{enumerate}[nosep, leftmargin=*, label=\arabic*.] 
    \item \textbf{Understand Issue \& Codebase}: Comprehend the problem from issue description. Examine codebase for context:
    \begin{enumerate}[nosep, leftmargin=*, label=(\arabic*)] 
        \item Code/patches referenced in issue.
        \item Unchanged/related parts of affected files.
    \end{enumerate}
    \item Analyze the Candidate Patches: For each patch, analyze its logic and intended fix. Consider whether the changes align with the issue description and coding conventions.
    \item verify its rationality with the rubric given below.
    \item The candidate patches have not yet applied to the repository, apply first before validate the patch
\end{enumerate}

\textbf{Rubric}
Your evaluation should focus on the following criteria:
\begin{verbatim}
${rubric}
\end{verbatim}

\textbf{Output Format:}
Reply in JSON:
\begin{verbatim}
{"result": "x" // id of the patch}
\end{verbatim}
\textbf{Analysis:} [Explain why Patch-x is correct.]

\textbf{Tasks:}
\begin{verbatim}
${task}
\end{verbatim}

\textbf{Candidate Patches}
\begin{verbatim}
${patches}
\end{verbatim}

\textbf{Output}
Now it's your turn.
\end{inputbox}

Rubric used during preference selection is as following
\begin{inputbox}{Rubric}
\small 
\textbf{1. Requirement Relevance}
\textbf{Definition:} How completely and precisely the patch satisfies **all** functional and non-functional requirements expressed or implied in the user’s task.

\rubrictable{
    \textbf{0} & \textbf{Severely Off-Topic} \\
    \textbf{1} & \textbf{Partial Coverage} \\
    \textbf{2} & \textbf{Highly Relevant} \\
    \textbf{3} & \textbf{Perfect Alignment}
}

\textbf{2. Code Accuracy}
\textbf{Definition:} Apply available tools to run the code and check for any compilation errors.

\rubrictable{
    \textbf{0} & \textbf{Broken} \\
    \textbf{1} & \textbf{Flawed} \\
    \textbf{2} & \textbf{Correct} \\
    \textbf{3} & \textbf{Robust \& Accurate}
}

\textbf{3. Change Precision}
\textbf{Definition:} How accurately the patch targets **only** the code that must change, avoiding extraneous edits.

\rubrictable{
    \textbf{0} & \textbf{Mis-targeted} \\
    \textbf{1} & \textbf{Imprecise} \\
    \textbf{2} & \textbf{Accurate} \\
    \textbf{3} & \textbf{Minimal \& Exact}
}

\textbf{4. Dependency \& Context Awareness}
\textbf{Definition:} Awareness of upstream/downstream dependencies and the completeness of associated updates (imports, call sites, configs, external contracts, backward compatibility).

\rubrictable{
    \textbf{0} & \textbf{Breaking Change} \\
    \textbf{1} & \textbf{Partial Awareness} \\
    \textbf{2} & \textbf{Internally Consistent} \\
    \textbf{3} & \textbf{System-Wide Vision}
}

\textbf{5. Code Quality}
\textbf{Definition:} Adherence to project style guides, language idioms, readability, and maintainability.

\rubrictable{
    \textbf{0} & \textbf{Poor} \\
    \textbf{1} & \textbf{Inconsistent Style} \\
    \textbf{2} & \textbf{Clean \& Comfortable} \\
    \textbf{3} & \textbf{Exemplary}
}

\textbf{6. Functionality Validation (Gating Criterion)}
\textbf{Definition:} Adherence to project style guides, language idioms, readability, and maintainability.

\rubrictable{
    \textbf{0} & \textbf{Any Failure} \\
    \textbf{3} & \textbf{Comprehensive \& Robust}
}
\end{inputbox}

A case study on the Preference Selector is as follows.
\begin{casebox}{Case Input: \$\{task\}}
    \small
    Modeling's separability\_matrix does not compute separability correctly for nested CompoundModels \\ 
    Consider the following model:
    ...
\end{casebox}

\begin{casebox}{Case Input: \$\{patches\}}
\small
\begin{lstlisting}[
basicstyle=\ttfamily\small,
columns=flexible,
breaklines=true,
tabsize=1, % 关键：设置缩进宽度为2个空格
showstringspaces=false
]
<patch id='A'>
diff --git a/astropy/modeling/separable.py b/astropy/modeling/separable.py
index a308e27297..45bea36085 100644
--- a/astropy/modeling/separable.py
+++ b/astropy/modeling/separable.py
@@ -242,7 +242,7 @@ def _cstack(left, right):
         cright = _coord_matrix(right, 'right', noutp)
     else:
         cright = np.zeros((noutp, right.shape[1]))
-        cright[-right.shape[0]:, -right.shape[1]:] = 1
+        cright[-right.shape[0]:, -right.shape[1]:] = right
 
     return np.hstack([cleft, cright])
...
</patch>
<patch id='B'>
...
</patch>
<patch id='C'>
...
</patch>
...
\end{lstlisting}
\end{casebox}

\begin{outputbox}{Case Agent Output}
\small
Perfect! All tests pass, including the new test that specifically covers the nested compound model issue.

\textbf{Summary}

I have successfully analyzed the issue and implemented the correct fix. Here's my analysis:

\textbf{Status: succeed}

\textbf{Result:}

\begin{verbatim}
{"result": "A"}
\end{verbatim}

\textbf{Analysis:}\\

Issue Understanding:
\begin{verbatim}
The issue was in the `_cstack` function in ...
\end{verbatim}

Root Cause:
...

\textbf{Why Patch A is correct:}

\begin{enumerate}
    \item \textbf{Core Fix}: All patches (A, B, C, D) contain the identical core fix to line 245
    
    \item \textbf{Test Coverage}: Patch A provides the most comprehensive test coverage, including:
    \begin{itemize}
        \item The exact issue scenario from the GitHub issue
        \item Edge cases with different nesting patterns
        \item Both left and right nested compound models
    \end{itemize}
    
    \item \textbf{Documentation}: Patch A includes clear docstrings explaining the test purpose
\end{enumerate}

...
\end{outputbox}

\end{document}